\documentclass{aa}
\usepackage{graphicx}
\usepackage{natbib}
\bibpunct{(}{)}{;}{a}{}{,}
\usepackage{psfig}

\newcommand{\be}{\begin{equation}}
\newcommand{\ee}{\end{equation}}
\newcommand{\bea}{\begin{eqnarray}}
\newcommand{\eea}{\end{eqnarray}}

\newcommand{\bfig}{\begin{figure}}
\newcommand{\efig}{\end{figure}}
\newcommand{\bfigw}{\begin{figure*}}
\newcommand{\efigw}{\end{figure*}}
\newcommand{\bmp}{\begin{minipage}}
\newcommand{\emp}{\end{minipage}}

\newcommand{\eg}{{e.g.}}
\newcommand{\ie}{{i.e. }}
\newcommand{\bpic}{\begin{picture}}
\newcommand{\epic}{\end{picture}}

\newcommand{\sct}{Sect.~}
\newcommand{\scts}{Sects.~}
\newcommand{\fg}{Fig.~}
\newcommand{\fgs}{Figs.~}
\newcommand{\eq}{Eq.~} 
\newcommand{\eqs}{Eqs.~} 
\newcommand{\ISO}{{\em ISO }}
\newcommand{\ISOc}{{\em ISO}}
\newcommand{\ISOPHOT}{{\em ISOPHOT }}

\newcommand{\ISOCAM}{{\em ISOCAM }}
\newcommand{\ISOCAMc}{{\em ISOCAM}}

\newcommand{\IRAS}{{\em IRAS }}

\newcommand{\Jy}{{\rm\thinspace Jy}}
\newcommand{\mJy}{{\rm\thinspace mJy}}

\newcommand{\m}{{\rm\thinspace m}}

\newcommand{\s}{{\rm\thinspace s}}

\newcommand{\mpsps}{\hbox{$\m\s^{-2}\,$}}

\newcommand{\um}{\hbox{$\mu {\rm m}$}}

\newcommand{\bband}{$B$--\,band~}

\newcommand{\rband}{$R$--\,band~}
\newcommand{\kband}{$K$--\,band~}
\newcommand{\iband}{$I$--\,band~}

\newcommand{\aips}{{\sc aips }}

\begin{document}
  \title{A 12$\um$ \ISOCAM Survey of the ESO-Sculptor Field
    \thanks{Based on observations collected at the European Southern 
      Observatory (ESO), La Silla, Chile, and on observations with \ISOc, 
      an ESA project with instruments 
      funded by ESA Member States (especially the PI countries: France, 
      Germany, the Netherlands and the United Kingdom) and with the 
      participation of ISAS and NASA.}}
  
  \subtitle{Data Reduction and Analysis\thanks{Table 4 is only available in 
      electronic form at the CDS via anonymous ftp to cdsarc.u-strasbg.fr 
      (130.79.128.5) or via http://cdsweb.u-strasbg.fr/cgi-bin/qcat?J/A+A/}}
  
 \titlerunning{12$\mu$m \ISOCAM of the ESS}
  
  \author{Nick Seymour\inst{1,2}, Brigitte Rocca-Volmerange\inst{1,3} and 
    Val\'erie de Lapparent\inst{1}}
  
  \authorrunning{Seymour, Rocca-Volmerange and de Lapparent}
  
  \offprints{Nick Seymour, email: seymour@ipac.caltech.edu}
  
  \institute{Institut d'Astrophysique de Paris, UMR7095 CNRS / Univ. Pierre \&    Marie Curie, 98 bis boulevard Arago, 75014 Paris, France.
    \and
    {\it Spitzer} Science Center, California Institute of Technology,
    Mail Code 220-6, 1200 East California Boulevard, Pasadena, CA 91125 USA.
    \and 
    Universit\'e Paris-Sud, B\^at. 121, F-91405 Orsay cedex, France.}
  
  \date{Received ; accepted 24/05/2007}

  \abstract{ 
    We present a detailed reduction of a mid-infrared 12$\,\mu$m (LW10 filter) 
    \ISOCAM open time observation performed on the ESO-Sculptor Survey field 
    \citep{arnouts97}. A complete catalogue of 142 sources (120 galaxies and 
    22 stars), detected with high significance (equivalent to 5$\sigma$), 
    is presented above an integrated flux density of 0.24$\mJy$. Star/galaxy 
    separation is performed by a detailed study of colour-colour diagrams. 
    The catalogue is complete 
    to 1$\mJy$ and below this flux density the incompleteness is corrected 
    using two independent methods. The first method uses stars and the second 
    uses optical counterparts of the {\it ISOCAM} galaxies;
    these methods yield consistent results. We also apply
    an empirical flux density calibration using stars in the field. For each 
    star, the 12\,$\mu$m flux density is derived by fitting optical colours 
    from a multi-band $\chi^2$ 
    to stellar templates  (BaSel-2.0) and using empirical optical-IR 
    colour-colour relations. This article is 
    a companion analysis to \citet{rocca07} where the $12\,\um$ faint galaxy 
    counts are presented and analysed per galaxy type with the evolutionary code P\'EGASE.3.
     }
  
  \maketitle
  
  \section{Introduction}
  
  Deep infrared surveys performed with \ISOCAM \citep{cesarsky96}
  aboard ESA's \ISO satellite \citep{kessler96} have greatly increased
  our knowledge of the faint IR background (\eg\space\citealp{aussel99}).  
  From dust emission, the
  mid-infrared (MIR) is an ideal wavelength domain to study
  the fundamental process of star formation at cosmological 
  distances. There is also evidence for strong evolution of sources in the 
  MIR including recent results from \ISO and {\it Spitzer} satellites 
  \citep{appleton04,pozzi04}. However, progress in this 
  area has been complicated by technical difficulties in reducing the 
  \ISOCAM data. Despite the recent advances of {\it Spitzer}, there is still 
  a great deal of information to be extracted from the \ISOCAM data.
    
  When examining faint sources with \ISOCAMc, to be confident in the
  reliability of a source, one must be sure to have removed all
  sources of flux variation above the background noise which are not
  due to astronomical objects. Principle amongst these are
  `glitches' caused by cosmic ray impacts on the SiGa
  detector. There is additional transient behavior
  comprising long term transients which are effectively slow
  variations in the background, and short term transients which occur
  when a pixel moves on and off a source causing an upward or downward
  transient respectively.  This memory effect is an unfortunate
  property of the type of detectors available at the time of
  \ISOCAMc's development. The temporal shape of this
  lagged response after a flux step has been corrected by a technique
  developed by \citet{abergel96} which successfully removes 
transients due to 
  strong sources or changes in illumination. Then the software developed by 
  \citet{starck99} known as PRETI (Pattern REcognition Technique for 
  \ISOCAM data) was designed to identify and remove the other artifacts 
  due to cosmic rays effects and residual low frequency variations.
  
  The ultimate flux density calibration of \ISOCAM has also been subject of 
  much research.
  Great care is not only needed to distinguish real astronomical
  sources from other transient effects in the detectors, but also in
  calculating the true flux density of these objects.  A detailed analysis of
  the behavior of the detector was applied to \ISO observations of
  the Hubble Deep Field (HDF) which used the LW2 ($6.75\um$) and LW3
  ($15\um$) filters \citep{aussel99} and was found to be successful at 
  removing most types of transients significantly above the noise. 
  The results of this analysis,
  although largely consistent at bright flux densities, are considerably
  different at fainter flux densities from those previously derived by
  \cite{serjeant97} from the same observations (there is however
  reasonable consistency with a further analysis of
  \citealt{desert99}). The method of Aussel et al. (1999) involved 
  PRETI to identify and correct the cosmic rays effects as well as to  
  removing the low frequency variation of the background left after applying
  the Abergel method; many simulations were
  performed to test the completeness and flux density calibration of these
  data.  Another approach, which is empirical and does not depend on
  simulations, is that of \citet{efstathiou00} who used stars to
  calibrate their \ISOPHOT $90\um$ survey. \citet{vaisanen02} and
  \citet{oliver02} also used stars to calibrate their \ISOCAM LW2 and 
  LW3 observations. Furthermore \citet{clements99} used 
  observations of stars to verify their flux density calibration.
  
  Over the last few years, the {\it Lari} method \citep{lari01} has 
  been successfully applied to other surveys 
  \citep{gruppioni02,pozzi03,vaccari05}. This technique involves a 
  full analysis of the history of each pixel and much simulation. 
  We did not use this method in our desire to avoid simulations, but 
  we note that several of these papers \citep{gruppioni02,vaccari05}, 
  as well as the more recent work of \cite{rodighiero04}, use the well known 
  IR properties of normal stars to verify their flux density calibration.

  Here, we present \ISOCAM observations which were designed to cover
  the portion of the ESO-Sculptor  faint galaxy redshift survey 
   (ESS; \citealt{arnouts97}) which
  is least affected by the cirrus confusion noise. {The \ISOCAM 
  area was selected using the IPAC/IRSKY software which measured a mean sky 
  flux density of 18.78 mJy/arcmin$^2$ at $12\,\um$ and a rms of 5.42 mJy/arcmin$^2$;
  these values were measured in 1.5 arcmin pixels in a large field of view 
  of $90'\times90'$ centered on the ESS field. 
  We also estimate the colour excess as $0.014<E_{B-V}<0.022$ on the survey area 
  \citep{schlegel97}.  The ESO-Sculptor Survey
  is located close to the South Galactic Pole, 
  and covers a strip of approximately $0.24\times1.53\deg=0.37$
  deg$^2$ with CCD photometry complete to Johnson-Cousins $B=24.5$,
  $V=24.0$ and $R_\mathrm{c}=23.5$.  The ESS also provides a nearly
  complete redshift sample to $R_\mathrm{c}=20.5$
  \citep{lapparent03a,lapparent04} over a sub-region of
  $0.24 \times 1.02\deg = 0.25$ deg$^2$. \ISOCAMc-ESS 
  thus provides a unique
  complement to the other existing surveys in its combination of i)
  $\sim700$ square arcminute sky area, ii) $BVR_\mathrm{c}$ CCD
  photometry and NIR (DENIS and 2MASS) photometry, and iii)
  spectroscopic completeness to $R_\mathrm{c}=20.5$.
  
  In the following, we present the data reduction and calibration of
  \ISOCAMc-ESS. The observations were 
  performed with the $\sim12\,\um$ LW10 \ISOCAM filter which was designed to 
  have a similar passband as the \IRAS $12\,\um$ band. This similarity allows us  
  to take advantage of the flux density calibration of the \IRAS data. 
  In \sct 2, we first
  describe the adopted procedure for source extraction (\sct 2.1) and
  astrometry (\sct 2.2). This analysis is followed by the flux density 
  re-calibration in
  the \sct 3. We provide the complete source catalogue in \sct 4. 
  
  \section{Observations and Data Reduction}
  \label{sec:dred}
  
  \begin{table*}
    \caption[]{\ISO observation log for 10 raster pointings: target
      name, coordinates, observation number (ION), calculated on-target
      time (CoTT), number of stabilizing exposures ($N_\mathrm{stab}$),
      and number of exposures ($N_\mathrm{exp}$).}
    \label{tab:fields}
    $$
    \begin{tabular}{@{}cccccccccccccc@{}}
      \hline
      \noalign{\smallskip}
      Target     &  RA (J2000)  &  Dec (J2000) & ION & CoTT & $N_\mathrm{stab}$  & $N_\mathrm{exp}$ \\
      & h\ \ m\ \ s  & $\deg$\ \ $'$\ \ $''$&     &      &           &         \\
      \noalign{\smallskip}
      \hline
      \noalign{\smallskip}
      SC\_LW10\_1 & 00 23 33.06  & -30 01 07.8  & 86  & 4832 & 10   & 13   \\
      SC\_LW10\_2 & 00 22 58.06  & -30 01 07.8  & 87  & 4832 & 10   & 13   \\
      SC\_LW10\_3 & 00 22 23.06  & -30 01 07.8  & 88  & 4832 & 10   & 13   \\
      SC\_LW10\_4 & 00 21 48.06  & -30 01 07.8  & 89  & 4832 & 10   & 13   \\
      SC\_LW10\_5 & 00 21 13.06  & -30 01 07.8  & 90  & 4832 & 10   & 13   \\
      SC\_LW10\_6 & 00 21 13.06  & -30 08 44.8  & 91  & 4832 & 10   & 13   \\
      SC\_LW10\_7 & 00 21 48.06  & -30 08 44.8  & 92  & 4832 & 10   & 13   \\
      SC\_LW10\_8 & 00 22 23.06  & -30 08 44.8  & 93  & 4832 & 10   & 13   \\
      SC\_LW10\_9 & 00 22 58.06  & -30 08 44.8  & 94  & 4832 & 10   & 13   \\
      SC\_LW10\_10 & 00 23 33.06  & -30 08 44.8  & 95  & 4512 & 10   & 12   \\
      \noalign{\smallskip}
      \hline
    \end{tabular}
    $$
    \end{table*}

    The \ISOCAM observations consist of 10 overlapping raster
    observations arranged in a $5\times2$ configuration centered on RA
    (J2000) 00$^\mathrm{h}$ 22$^\mathrm{m}$ 23.06$^\mathrm{s}$ and Dec
    (J2000) $-30^\circ$ 04$'$ 55.65$''$. Each of the 10 rasters (see
    Table~\ref{tab:fields}) is composed of $M\times N$ pointings 
    (with $M=N=8$) of the
    long wavelength (LW) detector of \ISOCAMc, each offset by $dM=dN=60''$
    along the axis of the detector. Table~\ref{tab:param} shows the
    parameters which were constant for all observations.  Each pointing
    of the $32\times32$ pixel detector used the $6''$ pixel field of
    view (PFOV) mirror so that the detector's view at each pointing was
    a $192''\times 192''$ area of the sky. The total field of view of the
    \ISOCAM survey is therefore approximately $0.3 \times 0.8$ deg$^2$\
    intersecting over $\sim80\%$ of the ESS spectroscopic area.  For
    \ISO observations it was not possible to request a particular
    orientation of the camera. In the event our observations were
    performed at $\sim45\deg$ to the axes of the $5\times2$ arrangement
    of the observations (see \fg\ref{fig:largemap}). 

    This arrangement left four very small patches of $\sim20''$ diameter 
    unobserved by
    ISOCAM, which are aligned at the middle declination of the ISO
    pointings Dec(J2000) $-30^\circ 04'50''$ and have the following
    values of RA(J2000): $0^\mathrm{h}23^\mathrm{m}15.4^\mathrm{s}$,
    $0^\mathrm{h}22^\mathrm{m}40.2^\mathrm{s}$,
    $0^\mathrm{h}22^\mathrm{m}5.2^\mathrm{s}$,
    $0^\mathrm{h}22^\mathrm{m}30.2^\mathrm{s}$ 
    (these 4 patches are barely visible in \fg\ref{fig:largemap}).

    \bfigw
    \centering
    \includegraphics[height=17cm,angle=270]{figures/snr.ps}
    \caption{{\bf [modified from journal version due to size of image]} 
      Signal-to-noise map of the combined ten \ISOCAM rasters. The two 
      horizontal lines mark the region covered by the ESS.}
    \label{fig:largemap}
    \efigw
    
    We used the latest version of
    the original raw data files obtained from the \ISO Data 
    Archive\footnote{http://www.iso.vilspa.esa.es/ida}, 
    corresponding to the 10 raster pointings with the most up-to-date 
    FITS header information. The data reduction was largely done with 
    the \ISOCAM Interactive Users Analysis System, called  CIA \citep{Ott97}.
    
    At this stage, PRETI \citep{starck99} was run within the CIA/IDL
    environment to remove the cosmic rays effects and residual variation 
    of the background, 
    after subtracting the transients, to perform the flat correction and to 
    suppress the time-varying baseline. We used the transient correction for 
    the on-source detector response as described in \citet{abergel96}. The 
    flux density of each pixel may be converted from ADUs to $\mJy$ using the 
    conversion given in CIA: 1 ADU$=0.242\mJy$ (more details are given in 
    \sct\ref{sec:fc} on the flux density calibration). The 64 
    independent pointings of each raster were then projected onto the 
    sky allowing for 
    the known distortion of the sky due to the optics of \ISOCAMc. 

    Although the original pixel size of the individual exposures was $6''$, 
    each raster image was finely re-sampled to $2''$ in order to increase 
    the accuracy of the sky projection.  Finally, the 10 raster pointing 
    images were combined on the sky plane producing an image map and a map 
    of the associated rms error at each pixel (henceforth referred to as 
    the ``noise'' map). The final pixel size of these maps, from the 
    re-projection of the ten raster images, was chosen to be $3''$ as a 
    compromise between retaining the accurate high resolution of the sky 
    projection and possible lightly over-sampling the PSF whose FWHM is of 
    order of the original $6''$ pixels. We note that some additional 
    correlated noise and potential small positional errors maybe introduced 
    by a second re-sampling of the image.

    \begin{table}
      \caption[]{\ISO observation parameters which are common to all rasters}
      \label{tab:param}
      $$
      \begin{tabular}{@{}cc@{}}
	\hline
	\noalign{\smallskip}
	Parameter     &  Value\\
	\noalign{\smallskip}
	\hline
	\noalign{\smallskip}
	Filter & LW10 \\
	Band Centre & 12\,$\um$ \\
	Gain & 2 \\
	$T_{int}$ & 5.04 s \\
	PFOV  & $6''$ \\
	M,N & 8,8 \\
	dM,dN & $60''$ \\
	\noalign{\smallskip}
	\hline
      \end{tabular}
      $$
    \end{table}
    
    \subsection{Source Extraction}

    \ISOCAM observations require specifically designed source extraction
    algorithms, as the noise in raster data is correlated and varies
    across the field. This effect is especially noticeable near the edge of a raster
    where there are fewer readouts per sky position; as a result, the
    border regions of the survey are noisier than the central part. One must
    therefore use the corresponding ``noise'' map to determine the
    significance level of a source and to avoid false detections on the border
    of the survey.
    
    In order to extract the \ISOCAM faint sources of the ESS field, we use
    the multi-scale vision model (MVM, \citealt{br95}) as applied to
    \ISOCAM data of the Hubble Deep Field by \citet{starck99}, and which
    is implemented in their {\it Multi-Resolution} (MR) software. This
    method searches for objects on different scales in wavelet space
    using the so called `\`a trous' algorithm (we refer the reader to
    \citealt{starck99} for full details).
    
    We apply the {\it Multi-Resolution} software with a detection
    threshold of $5\tau_w$ where $\tau_w$ is the noise level in wavelet
    space. $\tau_w$\ is not directly equivalent to the dispersion 
    of Gaussian data but gives a qualitative idea of the
    significance of our detection. The ultimate limit to the detection
    of sources is the rate at which false detections occur due to
    residual glitches in the noise which are too faint to be removed by
    PRETI. \citet{starck99} performed simulations to assess the
    reliability of their data and obtained a $2\%$ false detection rate
    at the completeness limit ($5\tau_w$).  Here we choose to perform the
    simple test of applying the source extraction to the negative of our
    sky image for different thresholds. We detect no sources in the
    negative images above $4.5\tau_w$. Hence we believe our detections
    to be quite robust, although we cannot quantify our false detection rate.
    We note, though, that we find optical counterparts to all
    our $12\,\um$ sources covered by the ESS within $6''$ and 
    with $R<25$ (see \sct 3). If randomly distributed we would have expected 
    only $30\%$ to have had optical counterparts within $6''$.

    Using the {\sc mrdetect} task from MR with a PSF model sampled at
    $3''$, and a detection threshold of $5\tau_w$, we search for
    objects down to the 4th wavelet scale and obtain 142 sources to a
    detection limit of $\sim 0.24\mJy$. This limit corresponds to the
    integrated flux density, reconstructed by the wavelet detection program,
    of the faintest object detected. The final source catalogue, after
    astrometry and flux density re-calibration, is presented in
    \sct\ref{sec:cat}.
    
    With the final goal of validating the adopted extraction method
    suited to raster data with ISO, we also investigate the application of 
    {\sc SExtractor} \citep{ba96} to our field. For a high,
    approximately equivalent detection threshold of $\sim5\sigma$, {\sc
    SExtractor} finds far fewer sources than MR. Of those sources
    found by {\sc SExtractor}, only $90\%$ are also found by MR. This
    percentage then decreases rapidly with lower thresholds (\ie $75\%$
    at $\sim3\sigma$).  Additionally, the consistent sources in both
    catalogues are found to have a random rms offset of $\sim1.5''$ in their
    position. These offsets are most likely due to the different strategies 
    of the two different codes, as MR searches for structure in wavelet space 
    and {\sc SExtractor} searches the standard sky-plane.
    When compared to the list of ESS optical sources, we
    find that the {\sc SExtractor} positions are marginally more
    accurate than those measured by MR (rms of $2.0''$ instead of
    $2.75''$). 
    This comparison thus confirms that a standard source extraction
    algorithm such as {\sc SExtractor} cannot compete with an extraction
    technique specifically designed for \ISOCAM data, such as MR.

    \subsection{Astrometry}
    
    A first check of the astrometry is obtained by cross-correlation of
    our \ISOCAM source list with the ESS bright objects
    ($R_{\rm c}<21$) located within $6''$. We obtain an rms offset
    of $2''$ with no systematic offset, thus indicating that the
    absolute astrometry for both \ISOCAM and ESS catalogues are
    reliable.  
    
    To obtain an independent astrometric calibration, we also searched
    for another infrared catalogue. The closest available data in
    wavelength is the 2 Micron All Sky Survey (2MASS) which includes
    $J$, $H$ and crucially \kband at $2\um$. As the 2MASS and \ISOCAM
    observations were taken within a few years of each other, the proper
    motion of stars (which make up most of the sources used for the astrometry)
    is unlikely to be a problem.  We cross-correlate our \ISOCAM list
    with the 2MASS Point Source Catalogue to search for objects within
    $3''$ of each other. This correlation yields 34 \ISOCAM objects with firm
    detections and provisional flux densities above $0.6\mJy$ which were also
    detected in the $K_s$-band. We then use these sources with the {\sc
      xtran} task in \aips to modify the header of the FITS file of the
    image and rms map.  This modification leads to a maximum change of $0.3''$ to the
    positions of the 142 final catalogue objects across the field and a 
    $0.1''$ improvement in the rms offset with respect to the ESS catalogue.

    \section{Flux Density Calibration}
    \label{sec:fc}
    
    \begin{table} \caption[]{Sub-sample of 13 stars selected for
	stellar fitting. The first column lists the source number from
	Table~\ref{tab:cat}. The next column contains the observed
	$I-\,$band magnitude from USNO (further photometry of these sources 
        is presented in table~\ref{tab:cat}). 
	The final 4 columns contain the
	results of the template fitting: the log surface gravity
	(\mpsps), the log metallicity ($Z/Z_0$), the effective
	temperature in Kelvin, and $\chi^2$ of the fit.}
      \label{tab:stars}
      $$
      \begin{tabular}{@{}rcccccccccccccc@{}}
	\hline
	\noalign{\smallskip}
	Source & $I_{USNO}$   &$\log(g)$&$\log(Z/Z_0)$&$T_\mathrm{eff}$& $\chi^2$ \\
	\noalign{\smallskip}
	\hline
	\noalign{\smallskip}
	1  & 10.260   & 3.00 & -1.7 & 6000 &  1.30 \\
	4  & 11.336   & 4.00 & -3.7 & 4500 &  0.66 \\
	5  & 10.601   & 3.00 & -2.0 & 6000 &  1.01 \\
	6  & 11.525   & 2.50 & -3.2 & 4500 &  1.48 \\
	18 & 11.898   & 4.00 & -4.0 & 5000 &  1.16 \\
	20 & 12.103   & 2.50 & -4.0 & 5500 &  1.68 \\
	34 & 12.594   & 5.00 & -2.7 & 4250 &  0.56 \\
	43 & 12.088   & 5.00 & -3.7 & 5250 &  0.44 \\
	46 & 12.338   & 5.00 & -0.7 & 5250 &  1.21 \\
	50 & 12.490   & 3.00 & -0.7 & 5500 &  0.51 \\
	53 & 12.917   & 5.00 & -2.2 & 4500 &  1.04 \\
	91 & 13.624   & 5.00 & -1.4 & 3750 &  1.04 \\
	130& 13.385   & 5.00 & -0.7 & 5750 &  1.10 \\
	\noalign{\smallskip}
	\hline
      \end{tabular}
      $$
    \end{table}

    Although selected to be far from the galactic plane, the relatively 
    large area of the ESS field provides a sample of stars of
    various types sufficient to estimate the empirical flux density
    calibration of the \ISOCAM observations. This areal size is a significant
    advantage compared to most other medium/deep \ISOCAM surveys (\eg\space
    the \ISO Hubble Deep Field North which covers only $\sim27$ square
    arcminutes of the sky and thus contains only a few calibrating
    stars\footnote{We note that the HDF South \citep{oliver02}, which
    is of a similar size, was calibrated using seven stars due to it's
    low galactic latitude.}).

    The following analysis uses a series of colour-colour diagrams and 
    relationships involving optical, NIR and $12\,\um$ colours. Due to the 
    high galactic latitude of our survey, the low extinction $E(B-V)\sim0.02$ 
    affects very little these colour-colour relations. For example, the $B$ to 
    \rband flux density ratio varies by less than $5\%$, and this percentage 
    is lower for flux density ratios of longer wavelength bands.

    In the following sub-sections, we describe the various stages of our
    calibration procedure. We first obtain for the detected \ISOCAM
    sources optical and NIR magnitudes by cross-identification with the
    ESS and various other existing catalogues (\sct \ref{sec:cross}).
    We then use colour-colour diagrams to identify stars among the
    \ISOCAM sources (\sct\ref{sec:star}).  The core of the calibration
    strategy uses a fitting procedure (D. Le Borgne, private communication) 
    to search for the best fit stellar template from the P\'EGASE
    library (\sct\ref{sec:fit}). We use the colours of the best fit 
    templates to predict $12\,\um$ flux densities from known IRAS 
    colour-colour relations \citealt[][(hereafter WCA)]{wca87} and then 
    convert to an ISO flux density.

    \subsection{Optical and NIR cross-identification}
    \label{sec:cross}
    
    The optical and NIR magnitudes are taken from a variety of sources:
    
    \begin{enumerate} 
    \item{Deep $B$, $V$ and $R_\mathrm{c}$-band data from ESS (including the 
      {\sc Sextractor} \citep{ba96} stellarity index - an indication of 
      how similar a 
      source is to the point-spread function, i.e. star-like or extended like 
      a galaxy} (see \citealp{arnouts97} for more details).
    \item{$B$, $R$ and \iband data from the USNO B catalogue (when $B$
      and \rband data are unavailable due to either saturation in the ESS,
      or masking of ESS diffraction spikes, or because the object is
      outside the ESS field). It should be noted that USNO
      magnitudes are $photographic$, but for convenience} we use them later 
      with the labels $B$, $R$ and \iband.  
    \item{\iband data from the DENIS Survey\footnote{http://www-denis.iap.fr/} 
       Extended Source Catalogue (ESC). We use the ESC as DENIS is deep enough 
       to detect more galaxies than stars.} 
    \item{$J$, $H$ and $K_s$-band from the 2MASS Point Source Catalogue 
      (PSC). We use the PSC as only the brighter \ISOCAM sources will be 
      detected and will mainly be stars.} 
    \end{enumerate}
    
    We first correlate the $12\,\um$ source list with the ESS catalogue
    using a $6''$ search radius which gives 111 objects out of 142
    with one or more potential optical counterparts. For sources 
    with one or 
    more optical counterpart we took the nearer the counterpart which in 11 
    out of 12 times was the brighter of two. The other source with two 
    counterparts seemed to be an interacting pair of equal brightness in the
    optical images.

    The remaining 31 objects are correlated with the USNO B catalogue in 
    a similar fashion. This correlation 
    leads to 23 further optical counterparts which include objects
    either outside the ESS field, or cut out of the ESS due to saturation
    or masking. This list of 134 optical sources is then correlated with
    the 2MASS data using the optical position and a $1''$ search radius,
    yielding 41 cross-identifications. \iband magnitudes from DENIS
    are also found for 79 sources with the same search radius. The more 
    accurate DENIS \iband is used in preference to the USNO 
    {\it photographic} \iband when available. There remains 7 objects 
    which are too faint to be detected in either USNO, 2MASS or DENIS 
    catalogues: sources 100 \& 110 are masked by diffraction spikes in the ESS
    and sources 12, 51, 81, 114 \& 137 are not coincident with the ESS 
    area. One source (source 44) is only detected in 2MASS.
    Hence all \ISOCAM sources covered by ESS 
    have optical counterparts with $R<25$. 
        
    \subsection{Star/galaxy separation and normal star selection}
    \label{sec:star}
     
    \subsubsection{$H-K_s$ versus $J-K_s$}
    We use the following colour/colour diagrams to separate stars from galaxies. 
    For all 41 objects with 2MASS counterpart, we select stars by
    examining their NIR colours using the criteria 
    
    \be
    \label{eq:bb1}
    H-K_s < 0.30
    \ee
    
    \be
    \label{eq:bb2}
    J-K_s < 1.0 
    \ee
    for normal stars from \cite{bb88}. This selection is illustrated in 
    \fg\ref{fig:ircol} and yields a separated sub-group of 22 objects.
    Note the ESS misidentified galaxy with $H-K_s\sim0.1$ in this plot 
    and others (object 13 in Table~\ref{tab:cat}). In \fg\ref{fig:ircol2}
    this object has a quite extreme position and is likely a very cool star 
    with a large MIR excess.

    \subsubsection{$H-K_s$ versus $K_s-[12]$} 
    
    To check for MIR excess due to dust and circumstellar material, we 
    examine $H-K_s$ versus $K_s-[12]$. The IRAS Explanatory
    Supplement\footnote{http://spider.ipac.caltech.edu/staff/tchester/exp.sup/}
    provides a zero magnitude flux density of $28.3\Jy$ for the 12\,$\um$
    magnitude. However, this value is not technically correct as it was obtained 
    by assuming that Vega is a blackbody from 10.6 $\um$ to 12\,$\um$, whereas 
    Vega has quite a large IR excess which affects both the 10.6 $\um$ flux 
    density and the spectrum. Here we adopt a value of $40.141\,\Jy$ from 
    \citet{cohen92}. Therefore:
    
    \begin{equation}
      \label{eq:iras}
	    [12]=-2.5\times\log\bigg(\frac{\mathrm F_{\Jy}}{40.141}\bigg)
    \end{equation}
    
    Among the 41 2MASS counterparts, 19 of the 22 stars selected previously 
    are in a cloud defined by 
    $0<H-K_s<0.3$ and $-0.5<K_s-[12]<0.5$ (see \fg\ref{fig:ircol1}) 
    and are unambiguously detected with a single optical/NIR counterpart 
    within $3''$. The other 5 objects with $H-K<0.4$, but $K-[12]>1.5$ 
    are stars from the group of 22 with MIR excess. 
    
    \subsubsection{$\log(f_{12}/f_R)$ versus $\log(f_B/f_R)$}
    
    To extend the analysis to objects too faint to be detected by
    2MASS (for which we have no NIR data) we examine the positions of
    all objects in a optical-MIR colour-colour diagram, shown
    in \fg\ref{fig:ircol2}. On this diagram is marked the loci of 
    blackbodies with a temperature ranging from 3000K (upper
    right) to $10^4$K (lower left). The 14 objects near the blackbody
    line are normal stars without IR excess; they define the 
    subsample that we analyse with the fitting procedure. All the
    14 objects belong to the subset of 19 objects with $K_s-[12]\sim0$
    and $H-K_s\sim0-0.25$ in \fg\ref{fig:ircol1}. The remaining 5
    objects are: i) 4 of the 5 {\sc Sextractor} stars identified in
    the ESS which lie immediately above the blackbody line and ii) the
    misidentified galaxy to the right of the graph with
    $\log(f_B/f_R)\sim3$. The fifth star in the row above the
    blackbody line (with $\log(f_B/f_R)=0.1$) has $K_s-[12]\sim1.5$ and
    hence does not satisfy the NIR/MIR colour selection criteria.
    Note that there are no objects previously not identified as stars
    in the lower part of \fg\ref{fig:ircol2}. 
    Therefore, among all the objects with optical counterparts, we
    identify from the colour-colour diagrams only 22 stars (i.e. all 
    and only those selected in \fg\ref{fig:ircol}). 
    With no further information available the 7 \ISOCAM sources without 
    either an optical or NIR counterpart are assumed to be galaxies.

    At this stage we change the DENIS $i-\,$band magnitudes from 
    the ESC to the PSC for all 22 sources found to be stars.

    \bfig
    \centering
    \includegraphics[height=8.5cm,angle=270]{figures/ir_colb2y.ps}
    \caption{NIR colour-colour diagram of 40/41 objects with 2MASS 
      counterparts (minus one objects with extreme $H-K$ colour, $\sim2$, 
      which lies off the plot). 
      The symbols refer to the stellarity index from the ESS 
      (see text): circles are galaxies, open stars are stars and crosses are 
      objects without counterpart in the ESS. The solid lines represent 
      the selection criteria of stars from \eqs\ref{eq:bb1} \& \ref{eq:bb2}.}
    \label{fig:ircol}
    \efig
    
    \bfig
    \centering
    \includegraphics[height=8.5cm,angle=270]{figures/ir_colb2x.ps}
    \caption{NIR-MIR colour-colour diagram of 40 objects with 2MASS 
      counterparts. The symbols are the same as in  \fg\ref{fig:ircol}}
    \label{fig:ircol1}
    \efig
    
    \bfig
    \centering
    \includegraphics[height=8.5cm,angle=270]{figures/ir_colz2b.ps}
    \caption{Optical-MIR colour-colour diagram of all objects with optical 
      (ESS/USNO) counterparts. The symbols are the same as in 
      \fgs\ref{fig:ircol} \& \ref{fig:ircol1}, but with the caveat that 
      $R$ and \bband magnitudes with no stellarity index are from the USNO
      catalogue (\ie the crosses). 
      The solid line represents the loci of blackbodies with a temperature 
      ranging from 3000K to 10000K for the Johnson-Cousins system. We note 
      this line would vary by $<0.1dex$ for the USNO filter system. 
      Furthermore, there are 7 \ISOCAM sources (three misidentified, open 
      stars, and four unknown, crosses) not in the ESS 
      with $\log f_{12}/f_R\sim10$ which are assumed to be galaxies.}
    \label{fig:ircol2}
    \efig
    
    \subsection{Template fitting}
    \label{sec:fit}
    
    For the template fitting we use the stellar library from the P\'EGASE.2  
    (http://www2.iap.fr/pegase) code. This  library  has a  significant
    coverage  of the HR  diagram. It  is based  on the  Kurucz library
    rearranged by \cite{lejeune97}: BaSel-2.0 (see Fioc \& 
    Rocca-Volmerange 1987, for details). The chi-squared fitting routine is 
    applied to the 14 objects of the selected subsample, all
    have $BRIJHK_s$ magnitudes with the $B$, $R$ and \iband magnitudes from
    the USNO catalogue, {\it not} the ESS catalogue, as they were all
    saturated or off the ESS field. 
    The 2MASS pass-bands come from the 2MASS
    web-page\footnote{www.ipac.caltech.edu/2mass/releases/allsky/index.html}. 
    The USNO pass-bands (from the Palomar Sky Survey and approximately 
    equivalent to Johnson $B$, $R$ and $I$--\,band) were taken from 
    \cite{reid91}. We remind the reader that the extinction is low enough 
    in this area of sky that its effect on the current analysis is small, 
    $\la5\%$ in the optical and less at longer wavelengths.
    
    The chi-squared fitting of the sample of 14 normal stars is carried out 
    in a 3 dimensional parameter
    space: effective temperature ($T_\mathrm{eff}$), surface gravity
    ($g$) and metallicity ($Z/Z_0$). The fits are constrained by the
    errors of each magnitude: 0.05 mag for 2MASS, $\sim0.05$ mag for
    DENIS, and 0.25 mag for USNO. The $\chi^2$ exhibit a sharp
    minimum as a function of $T_\mathrm{eff}$ and with a reasonable
    dependence on the surface gravity. For 13 out of the 14 stars,
    best-fit templates have $\chi^2 < 1.7$ (the 14th star has
    $\chi^2=4.0$ so we exclude it from our sample). 
    We show in \fg\ref{fig:fit} the
    stellar fit with the largest $\chi^2$ ($\chi^2 = 1.68$). The input
    magnitudes of the 13 stars are listed in Tables~\ref{tab:stars} and~\ref{tab:cat},
    along with the best fit values of the gravity, the metallicity, the
    effective temperature, and corresponding value of $\chi^2$. 
    
    Figure~\ref{fig:fit} illustrates that the NIR region of the SED is
    more constrained due the higher accuracies of the NIR magnitudes
    ($IJHK_s$).  This wavelength range corresponds to the region in which 
    the effective
    temperature is essentially defined, as it is the Wien-tail of the
    blackbody spectrum. From \fg\ref{fig:fit}, one might expect that
    there would be many other stellar templates consistent with this
    fit, but with a different SED below $0.8\um$. Although these different 
    optical SEDs would
    probably not change the effective temperature by much it could
    considerably affect the optical colours, $B-V$ and $R-I$, which the
    WCA colour-colour relations use (see \scts\ref{sec:pred}). To
    investigate this effect, we reran the fitting procedure to all 14
    stars without using the $B$ and \rband magnitudes (\ie without
    constraining the fit below \iband wavelengths). In all cases but
    one, the effective temperatures changes by only one step in
    parameter space ($200-250K$) and the SEDs below \iband are generally
    consistent with those derived using the $B$ and \rband magnitudes;
    similar values of $B-V$ and $R-I$ colours are also obtained.  The
    one case which has a considerably different effective temperature
    and SED is found to be almost degenerate with 2 sharp troughs in
    $\chi^2$ parameter space, the slightly higher of the $\chi^2$
    corresponding to the original 6 band fit; the other fit can easily
    be discarded as it is inconsistent with the observed $B$ and \rband
    magnitudes.  Therefore, we conclude that the large errors of the
    $B$ and \rband magnitudes do not significantly deteriorate the 
    quality of the fits.
    
    \bfig
    \centering
    \includegraphics[height=6.3cm,angle=0]{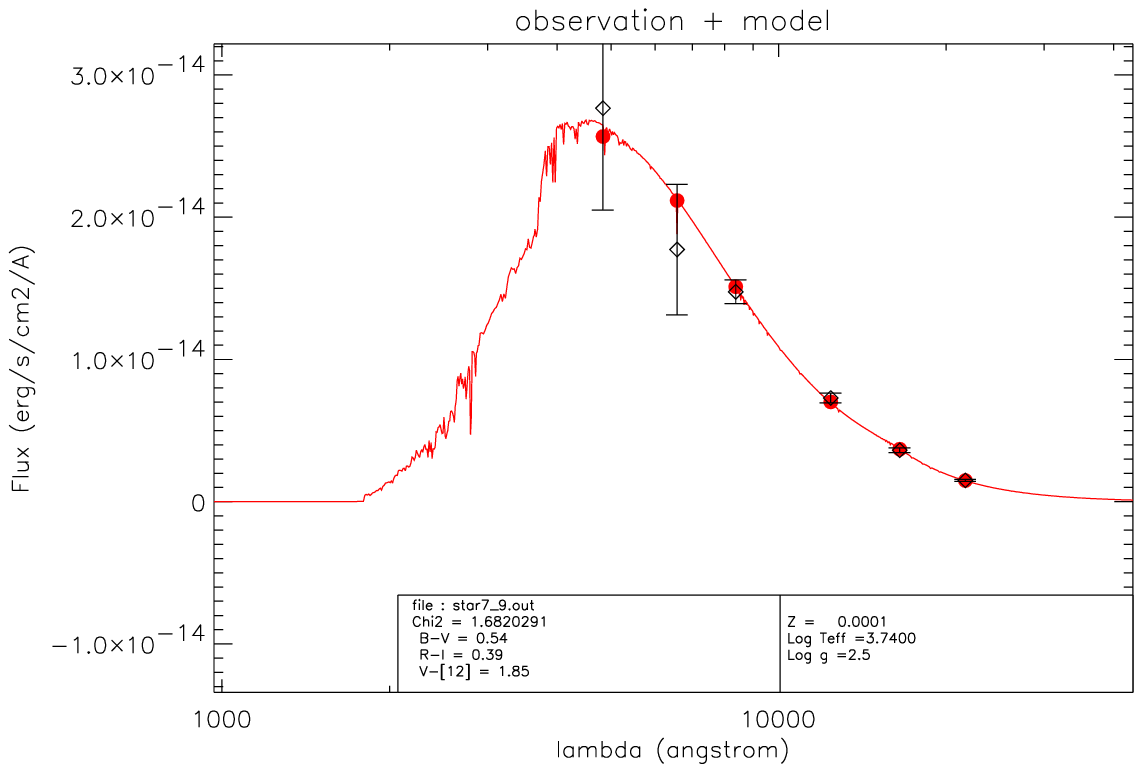}
    \caption{Result of the stellar fitting to \ISOCAM source 20 from 
      Table~\ref{tab:cat}. The open diamonds with error bars are the 
      observation magnitudes ($BRIJHK_s$ bands) whilst the solid circles are 
      the magnitudes of the best fit template (line) with $\chi^2 = 1.68$.
      Details of this fit are given in the panel below the plot.
      Although the fit is less constrained at shorter wavelengths due to
      the larger photometric errors in the USNO survey, it depends only weakly
      on this part of the spectrum (see text).}
    \label{fig:fit} 
    \efig

    \subsection{Predicted Flux}
    \label{sec:pred}
    
    Although the BaSeL-2.0 stellar libraries do extend to MIR 
    wavelengths they have not be widely tested, and then only at brighter 
    flux densities (\eg~ Cohen et al. 2003, using the Kurucz templates for
    certain A0-AV5 stars). We decide to use empirical stellar colours 
    to verify the calibration of our flux densities and use the 
    long-wavelength part of the stellar template as a consistency check.
    We use the two colour-colour relationships of WCA to derive 
    $12\,\um$ flux densities. These WCA relations are for IRAS $12\,\um$ flux 
    densities of stars which specifically relate the $V-[12]$ colour 
    with $B-V$ and $R-I$:

    \begin{eqnarray}
      \label{eq:col0}
      \begin{tabular}{@{}llll@{}}
	$V-[12]=0.05$ &$\!\!\!+~~3.13(B-V)$    &$-~~1.26(B-V)^2$ \\
                      &$\!\!\! +~~0.29(B-V)^3$ &$+~~0.16(B-V)^4$
      \end{tabular}
    \end{eqnarray}
    
    and
    
    \be
    \label{eq:col1}
    V-[12]=4.33(R-I)+0.14$  for $(R-I)<0.72
    \ee
    
    \be
    \label{eq:col2}
    V-[12]=2.69(R-I)+1.40$  for $(R-I)>0.72
    \ee
    
    These relationships were derived from \IRAS observations of bright normal 
    stars. We convert from magnitudes to flux density using the 40.141\,Jy 
    IRAS $12\,\um$ zero point of \citet{cohen92}. This zero point is very 
    similar to the value used in Cohen et al. (1987) who obtained $V-[12]$ 
    colours in agreement with the WCA colours. The \ISOCAM LW10 filter was 
    designed to have a
    similar wavelength-dependent response as the \IRAS $12\,\um$
    band\footnote{see http://www.iso.vilspa.esa.es/users/handbook/}, 
    but they are not identical. The ISO $12\,\um$ filter has a redder
    blue cutoff than the the IRAS $12\,\um$ filter. Using these 
    relationships we obtain predicted IRAS flux densities and which we can 
    then convert to ISO flux densities, by the ratios of the colour 
    corrections for the stars. The ratio of colour corrections, 
    $K_{12\um}^{IRAS}/K_{12\um}^{ISO}$, as defined in the respective 
    handbooks\footnote{IRAS: 
    {\tt irsa.ipac.caltech.edu/IRASdocs/exp.sup/ }
    and ISO: {\tt www.iso.esac.esa.int/manuals/HANDBOOK/}.}
    is computed by the authors to be 1.11 for stars of temperatures 2000-10000\,K.
    Hence we convert the IRAS to ISO flux densities by dividing
    the predicted WCA flux densities by this factor.    

    The $V$ magnitudes and $B-V$, $R-I$ colours which we use when
    applying \eqs\ref{eq:col0}, \ref{eq:col1} and \ref{eq:col2} to the
    13 stars listed in Table~\ref{tab:stars}, are those derived directly
    from best-fit template spectra. We apply for each star both
    relationships, thus yielding 2 estimates of the $12\,\um$ magnitude. 
    \eq~\ref{eq:iras} subsequently gives us 2 estimates of the flux density, 
    from which we derive an average flux density.  The ratio of this
    predicted \ISOCAM flux density to the initially measured \ISOCAM flux 
    density, i.e. the {\it observed} flux density, is plotted in 
    \fg\ref{fig:ratio}. We find a mean value of $1.01\pm0.17$ which 
    we also mark in \fg\ref{fig:ratio}. This mean value of the flux density 
    ratios includes the factor of 1.11 due to the colour corrections.
    The ratios presented in 
    \fg\ref{fig:ratio} indicate that, for our observational setup
    at least, the flux densities are systematically overestimated for the
    normal stars, with no apparent systematic variation in the offset
    value with $12\,\um$ flux density.  Due to the large PSF of these \ISOCAM
    observations, $6''$, slightly extended sources (galaxies) have a
    similar response to the detector as unresolved (stars) and hence we
    apply this correction to our entire catalogue (see
    Table~\ref{tab:cat}).
    Furthermore, the $1.01$ correction factor is not too dissimilar to the 
    combined flux density correction found by \cite{rodighiero04} 
    of 0.84 (from the combination of their projection bias, mosaic bias and
    stellar flux density correction: $0.84\times0.915\times1.097=0.84$).

    We are confident we have successfully excluded all stars which exhibit
    unusual properties (\eg~Be-stars or those with dust around
    them). As a test, we also calculate the ratio of predicted to observed 
    flux density using using the $12\,\um$ flux density derived from the Kurucz 
    stellar template. The ratio for the Kurucz template flux density is 
    $1.05\pm0.18$. We note that the Kurucz value is within good agreement 
    with the 
    value from our template fitting procedure, given the error bars,
    suggesting that the Kurucz models are not too far off at MIR wavelengths. 

    \bfig
    \centering
    \includegraphics[height=8.5cm,angle=270]{figures/correction3.ps}
    \caption{Ratio of predicted \ISOCAM $12\,\um$ flux density (from template 
      fitting and the relationships of WCA) to ``observed'' \ISOCAM 
      flux density plotted against observed $12\,\um$ magnitude. The mean ratio 
      is indicated by the horizontal solid line.}
    \label{fig:ratio}
    \efig
    
    \section{The $12\,\um$ Source Catalogue}
    \label{sec:cat}
    
    Table~\ref{tab:cat} lists the full flux density-calibrated $12\,\um$
    catalogue of 142 sources detected by \ISOCAM above the minimum
    integrated flux density of $0.24\mJy$. All
    available optical and NIR data are also given as well as
    classification (star/galaxy) and the source of the optical
    data. This catalogue is used to compute the faint galaxy
    counts at 12$\mu$m in the companion paper  \citep{rocca07}. A
    complementary paper on the luminosity function at 12\,$\mu$m is in
    preparation.
    
    The uncertainties in the 12\,$\mu$m flux density
    in Table~\ref{tab:cat} (column $[16]$) are obtained using: 

    \be
    \frac{\delta S_{12}}{S_{12}}\simeq\sqrt{\bigg(\frac{\mathrm dS}{\mathrm S}\bigg)^2
    +\bigg(\frac{0.047}{1.01}\bigg)^2} 
    \ee 
    The first term in the quadratic sum is the original relative
    uncertainty in flux density provided by the source count
    extraction with the MR software. The second term results from the
    uncertainty in the $1.01$ mean correction factor applied to the MR 
    flux density (see \sct\ref{sec:pred}): as the $1.01$
    factor is measured over 13 data points (see
    \fg\ref{fig:ratio}) with an rms dispersion of $0.17$ around the
    mean, we approximate its uncertainty to $0.17/\sqrt{13}=0.047$.
   
    Only the first 10 lines of the catalogue are presented here. The full 
    catalogue is available on the online version of this article.

    \begin{table*}
      \caption[]{First ten lines of the complete $12\,\um$ source catalogue. 
        The full catalogue will be available in the online version. The first 
        column contains the source number. The second column contains the 
        IAU designated naming with the prefix IES (ISO ESO-Sculptor). The third 
        and fourth columns contain the 
	RA and Dec (J2000). The fifth column indicates the classification of 
	each source (G=galaxy, S=star). The sixth column indicates the source 
	of the $BVR$ magnitudes (ESS = ESS, USNO = USNO survey, OFFF = 
	off ESS area and not detected by USNO, SPIK = hidden by a 
	diffraction spike in ESS and not detected by USNO and SATD = 
	saturated in ESS and not detected by USNO. The following 10 
        columns are, respectively the $BVRIJHK_s$ and the $12\,\um$ flux 
        density and its uncertainty in mJy. The $B_{U}$ \& $R_{U}$ magnitudes 
        from USNO, $BVR$ from ESS, \iband from DENIS and $JHK_s$ from 2MASS.}
      \label{tab:cat}
      \tiny
      \begin{tabular}{@{}rccccccccccccccc@{}}
	\hline
	\noalign{\smallskip}
	ID & name & RA (J2000)  & Dec (J2000) & S/G & U/E/S &$B_{U}$&$B$&$V$&$R_{U}$&$R$&$I$&$J$&$H$& $K_s$&$S_{12}$ \\
	\noalign{\smallskip}
	\hline
	\noalign{\smallskip}
 1 & IES J002337-295713 & 00 23 37.53 & -29 57 13.39 & S & USNO & 11.30 &       &       & 10.81 &       & 10.61 & 09.88 & 09.63 & 09.55 & $5.31\pm0.28$\\
 2 & IES J002350-300559 & 00 23 50.09 & -30 05 59.72 & G & ESS  &       & 23.24 & 22.86 &       & 22.34 &       &       &       &       & $4.95\pm0.30$\\
 3 & IES J002235-300430 & 00 22 35.59 & -30 04 30.96 & G & ESS  & 17.92 &       & 16.99 &       & 16.42 & 15.69 & 16.05 & 15.23 & 14.90 & $4.63\pm0.27$\\
 4 & IES J002234-301115 & 00 22 34.51 & -30 11 15.51 & S & USNO & 13.10 &       &       & 12.12 &       & 11.58 & 10.69 & 10.16 & 10.00 & $3.73\pm0.21$\\
 5 & IES J002333-301427 & 00 23 33.52 & -30 14 27.07 & S & USNO & 11.53 &       &       & 10.79 &       & 10.48 & 10.15 & 09.89 & 09.82 & $3.22\pm0.23$\\
 6 & IES J002057-295728 & 00 20 57.72 & -29 57 28.86 & S & USNO & 13.19 &       &       & 12.31 &       & 11.54 & 10.79 & 10.28 & 10.16 & $3.18\pm0.52$\\
 7 & IES J002305-300408 & 00 23 05.78 & -30 04 08.17 & G & ESS  &       & 17.46 & 16.36 &       & 15.82 &       & 15.73 & 15.05 & 14.67 & $3.14\pm0.19$\\
 8 & IES J002104-295913 & 00 21 04.34 & -29 59 13.49 & G & ESS  &       & 20.75 & 19.57 &       & 18.92 & 18.39 & 17.29 & 16.20 & 15.29 & $2.31\pm0.15$\\
 9 & IES J002302-300718 & 00 23 02.90 & -30 07 18.09 & G & ESS  &       & 19.49 & 18.67 &       & 18.27 & 17.95 &       &       &       & $2.24\pm0.15$\\
10 & IES J002113-301300 & 00 21 13.00 & -30 13 00.94 & G & ESS  & 20.42 &       &       &       & 18.98 & 18.45 &       &       &       & $2.08\pm0.15$\\
      \noalign{\smallskip}
      \hline
    \end{tabular}
  \end{table*}

  \section{Survey completeness}
  
  In order to use our \ISOCAM source catalogue to derive galaxy numbers 
  counts, one needs to evaluate the completeness of the catalogue as a 
  function of flux density.
  We use two independent empirical methods. Firstly, we use the same method as
  used in the previous section to determine the flux density of stars
  (\sct\ref{sec:fc}), which we extend to those stars not detected by \ISOCAMc. 
  The second method is based on the optical counterparts to galaxies
  associated with \ISOCAM sources, but detected with lower
  significance, in the interval $3\tau_w$ to $5\tau_w$.
  
  \subsection{Completeness from stars}
  \label{sec:comp}
  
  The method is based on the following stages: i) selecting stars from
  the 2MASS catalogue using their NIR colours, ii) confirming that
  they are stars by template fitting (as in \sct\ref{sec:fit}) and
  iii) using the results of the fitting to predict the $12\,\um$ flux
  density from the optical colours.
  
  We select stars from the 2MASS catalogue as all objects with $H-K<0.3$,
  $J-K<1$ and $K<14$ within the area of the \ISOCAM field. The two
  colour criteria are the same as were used earlier (\eqs\ref{eq:bb1} 
  \& \ref{eq:bb2}) and are typical of normal stars \citep{bb88,allen}. 
  The limiting magnitude criterion
  is intended to avoid selecting objects significantly fainter than the 
  detection limit of the \ISOCAM survey ($F_{12\um}\sim0.2\mJy$ is 
  approximately equivalent to $K\sim13$ mag), 
  but is faint enough to allow for stars with some MIR excess to be included. 
  The resulting catalogue contains
  51 stars. We discuss possible selection effects at the end of this
  section, especially the impact of stars with infrared excess.
  
  Using all available magnitudes from ESS, 2MASS, DENIS and USNO for each of
  the 51 stars, we find the best fit template spectrum from the
  library of \citet{lejeune97}, in the same fashion as in
  \sct\ref{sec:fit} for the 22 stars detected by \ISOCAM.
  We then use the optical-$12\,\um$ relations of WCA in \eqs\ref{eq:col1}
  and~\ref{eq:col2} to estimate the $12\,\um$ flux density. Three of the 51 stars
  which have predicted $F_{\rm 12\,\um}>0.2\mJy$ have fits with
  $3<\chi^2<10$. For these 3 stars, we use instead the relation:
  
  \be
  K-[12]=0.03\pm0.1 
  \label{eq:k12}
  \ee
  
  \noindent derived from the 13 well fitted stars of \sct\ref{sec:fit}.
  For all other stars, we find a well fit spectrum with low $\chi^2$ ($<3$). 
  
  Of the 51 stars, 18 (including the 3 stars with $\chi^2>3$) have
  predicted $12\,\um$ flux density above the detection limit, but were not found
  by the MR source extraction software. 
  This sample, including the 22 flux density calibrated, \ISOCAMc-detected 
  stars is then used to determine the fraction of 2MASS stars detected 
  by \ISOCAM at $12\,\um$ as a function of flux density. This result provides us 
  with a first estimate of the completeness, presented in \fg\ref{fig:comp2} 
  (red dotted line). The bins in \fg\ref{fig:comp2} have
  equal sizes in log flux density space and the errors for each bin are
  assumed to be Poisson.
  
  We now consider the possible selection effects. It is unlikely that
  we have missed stars due to the colour criteria (\eqs\ref{eq:bb1} 
  \& \ref{eq:bb2}) which are robust for normal stars, even those 
  with a MIR excess. It is also unlikely that we
  have included any galaxies as they would be poorly fitted by the
  stellar templates.  {\it But it is possible that some} of the selected
  stars have infrared excess and hence their true $12\,\um$ flux density is
  larger than that predicted. An excess of $K-[12] \sim 1-4$ would be
  equivalent to an increase of flux density equivalent to
  $\Delta\log(F_{\rm 12\,\mu m})\sim0.4-1.6$. Hence if our field contains stars
  with infrared excesses, then the completeness estimated here
  represents an upper limit. We note however that only 3 out of the 22
  stars detected by \ISOCAM have $12\,\um$ excesses. Therefore, we
  estimate that the completeness may be overestimated by 15\% at most,
  which is well inside the Poisson error bars. 
  
  Because the pixels are relatively large, $6''$, most galaxies 
  appear unresolved to \ISOCAMc.  We can then assume that the
  sensitivity of the detector is the same for unresolved objects,
  \ie stars, as for resolved objects like galaxies. Hence the
  completeness in galaxies is likely similar to that of stars and can
  be used as such. If the completeness for galaxies is however affected by
  the non-detection of low surface brightness objects, then again the
  estimate plotted in \fg\ref{fig:comp2} is an upper limit. We also
  suspect that this effect is smaller than the plotted Poisson errors.
  
  \subsection{Completeness from low significance sources}
  \label{sec:weak}
  
  The second method provides an independent correction to the source
  counts.  It is based on the association of \ISOCAM sources detected
  at low significance with optically detected galaxies. A priori, some
  of the \ISOCAM sources detected with a detection threshold located
  in the interval $3\tau_w$ to $5\tau_w$ may be real. By examining
  their association with the ESS optical sources, we can evaluate
  their reality in a statistical way. We use PRETI as in 
  \sct\ref{sec:dred}, but with a
  detection threshold in wavelet space of $3\tau_w$. This threshold leads to
  detection of 328 sources, to be compared with the $5\tau_w$ list of
  142 objects presented in Table~\ref{tab:cat}. After
  correcting the flux density of each object by the correction found in
  section~\ref{sec:pred}, and removing the sources with a flux density below
  the detection limit of $0.24\mJy$, we end up with 292 potential
  sources (\ie the 142 sources of the original catalogue and 150 new 
  sources of lower significance). 
  
  We then cross-correlate the 150 new $3\tau_w$ \ISOCAM
  sources with the ESS catalogue. Using a $3''$ ($6''$) search radius
  we find 74 (110) optical counterparts. From $f_{12}/f_R$ versus $f_B/f_R$ 
  diagrams, like \fg\ref{fig:ircol2}, all these sources are found to 
  be galaxies. As some of
  these matches are by chance, due to the size of our search radius
  and the space density of optical sources, we evaluate the excess
  matches by offsetting the positions of the 2 data sets by increasing
  multiples of half an arcsecond in RA and Dec. At large offsets
  ($\ga6''$), the number of matches becomes roughly constant with a
  value around 20 (55) for the $3''$ ($6''$) search radius. These numbers are
  comparable with the theoretical numbers of 17 (70) sources expected
  within the $3''$ ($6''$) search radius if one uses the sky density
  of objects and assumes a random distribution. 
  
  We now assume that adding the sources from either the $3''$ or $6''$
  list to the high-significance list of 142 sources, yields a nearly
  `complete' source catalogue. This assumption is reasonable because
  there is a large excess of \ISOCAM coincidences with the ESS above
  the theoretical and empirical random values. A `real' object,
  detected with $[12]<13$ (corresponding to the $12\,\um$ detection
  limit), must have $V<22.5$ if it is a star-forming galaxy at $z\la1$. 
  This selection is derived from typical, maximum $K_s-[12]$ colours ($\sim5$, 
  see \fg\ref{fig:ircol1}) and 
  typical, maximum $V-K_s$ colour ($\sim4.5$). The possibility
  that some sources are not real is taken into account in the errors 
  on the source counts which are derived in \citet{rocca07}.

  The resulting completeness, defined as the ratio of the uncorrected
  number counts from the $5\tau_w$ list to the counts from the `complete' 
  list, is plotted as a function of flux density in \fg\ref{fig:comp2} 
  (solid line). The bins again have equal sizes in log flux 
  density space, albeit smaller than before, and the errors
  are simply Poisson. Note that the curve corresponding to a search
  radius of $3''$ is indistinguishable from that with $6''$ plotted in
  \fg\ref{fig:comp2}. 

  The sources counts extended to $3\tau_w$ are subject to several
  biases: they probably include some sources which are false and might
  still miss some real sources which would be above our $0.24\mJy$ detection
  threshold. The former would lead to an over-estimation of the source
  counts (although we include it in our error), and the
  latter, to an under-estimation.  It is nevertheless significant that
  the source counts from both the low-significance lists (with $3''$
  and $6''$ search radii) yield a consistent incompleteness with that
  derived from stars in the previous section (see \fg\ref{fig:comp2}).
  Although difficult to quantify, the agreement of these 2 independent
  methods for estimating the completeness indicates that these 2 
  selection effects affecting the low-significance sources cancel out
  to a certain extent. 

  Therefore both the full and dotted lines on \fgs\ref{fig:comp2} show 
  that our \ISOCAM catalogue is complete to $\sim1$mJy, with a linearly
  decreasing completeness in log flux density down to the our flux 
  density detection
  limit of $0.24$mJy.  A complete analysis of the \ISOCAMc-ESS
  galaxy number-counts along with the fitting of theoretical models
  are presented in the companion paper \citep{rocca07}.
  
  \bfig
  \centerline{\psfig{figure=figures/completeness_gal.ps,height=6.3cm,angle=270}}
  \caption{Completeness of the \ISOCAM catalogue as derived from both methods:
    stellar template fitting to the 2MASS stars in the ESS field and the 
    inclusion of lower significance \ISOCAM sources with a 
    detection threshold between $3\tau_w$ and $5\tau_w$ and ESO-Sculptor 
    optical counterparts detected within $6''$. The dotted red line shows 
    the fraction of detected stars as a function of predicted flux density, 
    with Poisson errors. The solid line shows the fraction of
    galaxies detected with a high significance ($5\tau_w$) from all
    galaxies detected with a lower significance ($3\tau_w$) as a function 
    of predicted flux density, with Poisson errors.}  
  \label{fig:comp2} 
  \efig
  
  \section{Conclusions}
  
  We present the data reduction of \ISOCAM observations performed with
  the LW10 filter centered near $12\,\um$, mostly in the field of the optical
  ESO-Sculptor survey \citep{arnouts97,lapparent03a}.  The data reduction of the
  \ISOCAM rasters is performed as by \citet{aussel99}: we use the
  multi-scale vision model of \citet{br95} implemented into the MR
  software \citep{starck99}, along with the PRETI algorithm aimed at
  removing all image artifacts above the noise level.  Using a
  detection threshold of $5\tau_w$ where $\tau_w$ is the noise level 
  in wavelet space, we reach a detection limit of $\sim 0.24\mJy$. 
  
  The final catalogue contains 142 \ISOCAM sources with optical
  counterparts in the ESO-Sculptor survey, which we complement by
  optical and NIR magnitudes from the USNO B catalogue, and the 2MASS
  and DENIS surveys.  Optical, near-infrared and  mid-infrared 
  colour-colour diagrams
  subsequently allow us to identify 22 sources as stars and 120 as
  galaxies, which dominate at faint flux densities. By template fitting of 13
  of these stars, we derive their predicted $12\,\um$ flux density using the
  template optical colours combined with the optical-mid-infrared colour
  relations of \citet{wca87}. By comparison with the observed \ISOCAM
  flux densities normalized using the \IRAS $12\,\um$ zero-point, we measure
  that the observed flux densities systematically overestimate the theoretical
  flux densities by a factor of 1.16 (1/0.86). We use this offset to correct 
  all flux densities (\ie stars
  and galaxies).  We use a similar method (predicting the $12\,\um$ flux density
  of stars) to determine the completeness of our survey as a function
  of flux density. This completeness function is found to be in good agreement
  with that from a statistical study of the coincidence of low
  significance \ISOCAM sources with the ESO-Sculptor optical sources.
  
  The \ISOCAMc-ESO-Sculptor-Survey catalogue obtained here is used to analyse
  the mid-infrared galaxy number counts in \citet{rocca07}. By using
  the available ESO-Sculptor redshifts \citep{lapparent03a}, we will also
  derive the $12\,\um$ luminosity function, which is a valuable tool for
  interpreting the deep mid-infrared source counts and performing a
  detailed study of the evolution of galaxies at $12\,\um$, a wavelength
  range unavailable to the {\it MIPS} and {\it IRAC } instruments
  aboard {it Spitzer} \citep{gallagher03}.
  
  \begin{acknowledgements}
    We thank the referee for the many constructive comments improving the 
    quality of this paper.
    We give much thanks to Ren\'e Gastaud, Jean-Luc Starck, David Elbaz 
    (SAp/CEA), Carlos del Burgo (Heidelberg) and Emmanuel Bertin (IAP).
    We are also grateful to Damien Le Borgne (CEA) 
    for aiding us in the use of his software. We thank Herv\'e Aussel for 
    useful discussions. Part of this work (NS) was supported by the
    \emph{Probing\- the Origin\- of the\- Extragalactic\- background
      (POE)\/}, European Network number HPRN-CT-2000-00138.  This
    publication makes use of data products from the Two Micron All Sky
    Survey, which is a joint project of the University of
    Massachusetts and the Infrared Processing and Analysis
    Center/California Institute of Technology, funded by the National
    Aeronautics and Space Administration and the National Science
    Foundation. Additionally this research uses the USNOFS Image and
    Catalogue Archive operated by the United States Naval Observatory,
    Flagstaff Station (http://www.nofs.navy.mil/data/fchpix/). We also
    thank Gary Mamon for kind permission to use data from the DENIS
    survey. The DENIS project has been partly funded by the SCIENCE 
    and the HCM plans of the European Commission under grants CT920791 
    and CT940627. It is supported by INSU, MEN and CNRS in France, by 
    the State of Baden-W\"u rttemberg in Germany, by DGICYT in Spain, by CNR 
    in Italy, by FFwFBWF in Austria, by FAPESP in Brazil, by OTKA grants 
    F-4239 and F-013990 in Hungary, and by the ESO C\&EE grant A-04-046.
    The ISOCAM data presented in this paper were analysed using 
    `CIA', a joint development by the ESA Astrophysics Division and the 
    ISOCAM Consortium. The ISOCAM Consortium is led by the ISOCAM PI, 
    C. Cesarsky.
    
  \end{acknowledgements}
  
  \bibliographystyle{aa}

\end{document}